\documentclass[reprint,nofootinbib,amsmath,amssymb,aps,floatfix,superscriptaddress,twocolumn]{revtex4-2}
\usepackage{graphicx}
\usepackage{dcolumn}
\usepackage[colorlinks,linkcolor=magenta,anchorcolor=cyan,citecolor=blue]{hyperref}
\usepackage{bm}
\usepackage[multiple]{footmisc}

\begin{document}
\title{Energy extraction via Comisso-Asenjo mechanism from rotating hairy black hole }

\author{Zhen Li} 
\email{zhen.li@nbi.ku.dk}
\affiliation{DARK, Niels Bohr Institute, University of Copenhagen, Jagtvej 128, 2200 Copenhagen Ø, Denmark}
\author{Faqiang Yuan}
\email{yfq@mail.bnu.edu.cn}
\affiliation{Department of Physics, Beijing Normal University, Beijing, 100875, China}

\date{\today}

\begin{abstract}
It was demonstrated by Comisso and Asenjo that the magnetic reconnection in the ergosphere is a promising mechanism to extract energy from the rotating Kerr black hole. In this work, we investigate the role of Comisso-Asenjo mechanism in energy extraction from the newly suggested rotating hairy black holes which have an extra hair due to the additional surrounding sources, such as dark matter or dark energy. We examine how the hairy parameters characterized the hair affect the magnetic reconnection process in addition to other important variables of the Comisso-Asenjo process, including the parameter spaces that permit energy extraction, the power, efficiency and power ratios with respect to the Blandford-Znajek mechanism. 
\end{abstract}
\maketitle

\section{Introduction}

It is widely accepted that the physically rotating black holes in the universe are described by the Kerr metric, which are characterized by two parameters, i.e., the mass and the spin. The famous No Hair theorem in classical general relativity \cite{nh1,nh2,nh3,nh4,nh5} also states that the Kerr metric has no other 'hair' besides the mass and spin. However, the unknown dark matter and dark energy, which may originate from new fundamental fields, could be an additional source of the black hole such that Kerr black holes obtain a new hair and deviate from the Kerr metric \cite{dk}. Based on this novel idea, recently, the hairy black hole and later its rotating counterpart were obtained using the gravitational decoupling approach (GD) \cite{gd0,gd} which has been created expressly to describe the deformation of known solutions of general relativity brought on by additional sources \cite{gd1,gd2}. There are numerous theoretical and observational investigations on this (rotating) hairy black hole \cite{inv1,inv2,inv3,inv31,inv4,inv5,inv6,inv7}.

On the other hand, a rapidly rotating black hole will produce an antiparallel magnetic field configuration in the equatorial plane \cite{mr1,mr2}. Comisso and Asenjo has shown that, when the aspect ratio of the current sheet approaches a critical value \cite{mr7,mr8,mr9}, this distinctive magnetic field configuration could result in a quick magnetic reconnection process inside the ergosphere, which is capable of converting a large amount of magnetic energy into the kinematic energy of the plasma particles which escape from the black hole to infinity \cite{mr11}. In order to distinguish this new mechanism from earlier attempt \cite{mr10} at magnetic reconnection, we will refer to it as the Comisso-Asenjo mechanism. Numerous numerical simulations have also demonstrated that the Comisso-Asenjo reconnection process always occurs at a dominant point \cite{mr3,mr5,mr6}. The frame dragging effect of a rotating black hole causes this process to occur again and again. Such a mechanism might be more effective than others for extracting the rotational energy from black holes, such as the Penrose process \cite{p1,p2} and the Blandford-Znajek mechanism \cite{bz4}. The Kerr black hole was the first black hole to be studied with energy extraction via Comisso-Asenjo mechanism \cite{mr11} and most lately applied to the other rotating black holes \cite{mrr1,mrr2,mrr3,mrr4,mrr5,mrr6}.

In this work, we look into the role of the Comisso-Asenjo mechanism in the energy extraction from the rotating hairy black hole. Since this process may result in more high energy astrophysical phenomena, therefore, this work may provide us phenomenological insights on testing the rotating hairy black hole as well as the No Hair theorem over a larger spectrum of future observations.

The structure of this paper is as follows: The rotating hairy black hole spacetime is presented in Sec.\ref{sec2}. Then in Sec.\ref{sec3}, The Comisso-Asenjo mechanism formulas will be discussed, along with the equations for plasma energy at infinity density per enthalpy and the conditions for energy extraction. Next in Sec.\ref{sec4}, use the Comisso-Asenjo mechanism formulas in the previous section, We will investigate the parameter spaces of energy extraction via magnetic reconnection. In Sec.\ref{sec5}, We study the power and efficiency of this mechanism using various parameter combinations. In this section, we also examine the power ratios of Comisso-Asenjo mechanism comparing to the Blandford-Znajek mechanism. At last, we will make a conclusion in Section.\ref{sec6}.

\section{Rotating hairy black hole}\label{sec2}
The GD method, which is specifically created to find deformation of the known solution of GR \cite{gd1,gd2}, was used to derive a rotating hairy black hole in \cite{gd}. Its line element square can be read as follows in the Boyer-Lindquist coordinates:
\begin{equation}\label{metric}
d s^{2}=g_{t t} d t^{2}+g_{r r} d r^{2}+g_{\theta \theta} d \theta^{2}+g_{\phi \phi} d \phi^{2}+2 g_{t \phi} d t d \phi
\end{equation}
where the metric components are given by 
\begin{align}
g_{t t}&= -\left[\frac{\Delta-a^{2} \sin ^{2} \theta}{\Sigma}\right] \quad \quad\quad\quad\quad \quad\,\,\,
g_{r r} =\frac{\Sigma}{\Delta} \nonumber\\
g_{t\phi}&=- a \sin ^{2} \theta\left[1-\frac{\Delta-a^{2} \sin ^{2} \theta}{\Sigma}\right]\quad\quad g_{\theta \theta} =\Sigma \nonumber\\
g_{\phi\phi}&=\sin ^{2} \theta\left[\Sigma+a^{2} \sin ^{2} \theta\left(2-\frac{\Delta-a^{2} \sin ^{2} \theta}{\Sigma}\right)\right]
\end{align}
with $\Delta=r^{2}+a^{2}-2 M r+\lambda r^{2} e^{-r /\left(M-\frac{h_{0}}{2}\right)}$, and $\Sigma=r^{2}+a^{2} \cos ^{2} \theta$. $M$, $a$ denote the black hole mass and spin. The primary hair $h_0$ measures the rise in entropy generated by the hair and must meet the requirement $h_0 \le 2M \equiv h_K$ in order to achieve asymptotic flatness. $\lambda$ measures deviation from the standard Kerr black holes and is related to $h_0$ via the equation $h_0 = \lambda h$. The Kerr metric, which denotes the absence of surrounding matter, is what this spacetime reduces to when $\lambda=0$. 

We will get horizons from below equation
\begin{equation}\label{eh}
\Delta=r^{2}+a^{2}-2 M r+\lambda r^{2} e^{-r /\left(M-\frac{h_{0}}{2}\right)}=0
\end{equation}
analytical solutions, however, do not exist. In \cite{inv2}, it was discussed how different hairy parameters affected the numerical results of the horizon structure and the ergosphere. We will restrict our discussion on the regime of hairy parameters such that (\ref{eh}) has two distinct real solutions (see\cite{inv2}).

\section{Energy extraction via Comisso-Asenjo mechanism }\label{sec3}
During the Comisso-Asenjo magnetic reconnection process, one part of the plasma or flux is accelerated while the other part decelerates in the opposite direction. The energy is taken from the rotating black hole by magnetic reconnection if the decelerated part has a negative energy and the accelerated part has an energy greater than its rest mass and thermal energy at infinity \cite{mr11}.
We now give the formulas of Comisso-Asenjo mechanism for figuring the energy at infinity related to accelerated and decelerated plasma in the rotating hairy black hole spacetime. In the so-called "zero-angular-momentum-observer" (ZAMO) frame \cite{zamo}, it is more convenient to assess the plasma energy density. The metric (\ref{metric}) in ZAMO frame is given by
\begin{equation}
d s^{2}=-d \hat{t}^{2}+\sum_{i=1}^{3}\left(d \hat{x}^{i}\right)^{2}=\eta_{\mu \nu} d \hat{x}^{\mu} d \hat{x}^{\nu}
\end{equation}
which is a Minkowski metric, where
\begin{equation}
d \hat{t}=\alpha d t, \quad d \hat{x}^{i}=\sqrt{g_{i i}} d x^{i}-\alpha \beta^{i} d t
\end{equation}
and 
\begin{equation}
\alpha=\left(-g_{t t}+\frac{g_{\phi t}^{2}}{g_{\phi \phi}}\right)^{1 / 2},\quad \beta^{\phi}=\frac{\sqrt{g_{\phi \phi}} \omega^{\phi}}{\alpha}
\end{equation}
we define $\omega^{\phi}=-g_{\phi t} / g_{\phi \phi}$ as the angular velocity of the frame dragging due to the rotating hairy spacetime.

The one fluid approximation energy-momentum tensor of this system, in Boyer-Lindquist coordinates, takes form as
\begin{equation}
T^{\mu \nu}=pg^{\mu \nu}+w U^{\mu}U^{\nu}+F^{\mu}_{\,\,\,\delta}F^{\nu \delta}-\frac{1}{4}g^{\mu \nu}F^{\rho \delta}F_{\rho \delta}
\end{equation}
where $p$, $w$, $U^{\mu}$, and $F_{\mu \nu}$ are respectively the proper plasma pressure, enthalpy density, four-velocity, and electromagnetic field tensor. With this energy-momentum tensor, we can get the so-called “energy-at-infinity” density $e^{\infty}=-\alpha g_{\mu 0}T^{\mu 0}=e^{\infty}_{hyd}+e^{\infty}_{em}$ \cite{mr10}, where $e^{\infty}_{hyd}$ and $e^{\infty}_{em}$ are respectively the hydrodynamic energy-at-infinity density and the electromagnetic energy-at-infinity density, and they are given by
\begin{align}
e^{\infty}_{hyd}&=\alpha(w \hat \gamma^2-p)+\alpha\beta^{\phi}w\hat \gamma^2 \hat v^{\phi}\nonumber\\
e^{\infty}_{em}&=\frac{\alpha}{2} (\hat B^2+\hat E^2)+(\hat{\bf B}\times \hat{\bf E})_{\phi}
\end{align}
where $\hat \gamma=\hat U^0=1/\sqrt{1-{\textstyle \sum_{i=1}^{3}} (d \hat v^i)^2}$, $\hat B^i=\epsilon ^{ijk}\hat F_{jk}/2$, $\hat E^i=\hat F_{i0}$ are the Lorentz factor, the components of magnetic and electric fields respectively. $v^{\phi}$ denotes the azimuthal component of the plasma outflow velocity in the ZAMO frame. Given the transformation of vectors between the Boyer-Lindquist frame and the ZAMO frame discussed in \cite{mr10,mr11}, the hat above these quantities indicates that we assess the quantity in the ZAMO frame.

Considering that almost all the magnetic energy is converted into kinetic energy of plasma during the Comisso-Asenjo magnetic reconnection process, as well as assuming that the plasma element is incompressible and adiabatic, we have \cite{mr10}
\begin{equation}\label{a1}
e^{\infty}= \alpha\left(w(\hat \gamma +\beta^{\phi}\hat \gamma\hat v^{\phi})+\frac{p}{\hat \gamma}\right)
\end{equation}

The local rest frame of plasma during the magnetic reconnection process, $\bar x^{\mu}=(\bar x^{0}, \bar x^{1}, \bar x^{2}, \bar x^{3})$ with $\bar x^{1}= r$ and $ \bar x^{3}=\phi$, rotates with Keplerian angular velocity $\Omega_K=d \phi/{d t}$ in the equatorial plane from the perspective of a Boyer-Lindquist observer, and also rotates with Keplerian angular velocity $\hat v^K$ in the ZAMO frame which is given by 
\begin{equation} \label{v1}
\hat v^K= \frac{\Omega_K\sqrt{g_{\phi\phi}}}{\alpha}-\beta^{\phi}
\end{equation}
We use the symbol $\bar v_{out}$ to represent the outflow velocity measured in the local rest frame. $\bar v_{out}$ is related to the properties of plasma magnetization and it could be expressed as \cite{mr11}
\begin{equation} \label{v2}
\bar v_{out}=\sqrt{\frac{\sigma_0}{\sigma_0+1}}
\end{equation}
where $\sigma_0= B_0^2/w_0$ is the plasma magnetization upstream of the reconnection layer, $B_0$ is the asymptotic macro-scale magnetic field and $w_0$ is the enthalpy density of the plasma.
The outflow velocity measured in the ZAMO frame is
\begin{equation}\label{v3}
\hat v^{\phi}_{\pm}=\frac{\hat v^K\pm \bar v_{out}cos(\xi) }{1\pm \hat v_K \bar v_{out}cos(\xi)}
\end{equation}
where $\pm$ denotes the outflow velocity with respect to the black hole's rotation and has corotating (+) and counterrotating (-) directions. They also represent, respectively, the accelerated and decelerated portions of the plasma. $\xi=arctan(\bar v^1/\bar v^3)$ is the plasma orientation angle, $\bar v^1$ and $\bar v^3$ are the radial and azimuthal components of plasma velocities in the local rest frame.

With the above Eq.(\ref{a1}) and Eq.(\ref{v1}), (\ref{v2}),(\ref{v3}), then the plasma energy-at-infinity density per enthalpy as $\epsilon_{\pm}^{\infty}=e_{\pm}^{\infty}/w$ becomes \cite{mr11}
\begin{align}\label{ep}
\epsilon_{ \pm}^{\infty}=\alpha \hat{\gamma}_{K}[& \left(1+\beta^{\phi} \hat{v}_{K}\right)\left(1+\sigma_{0}\right)^{1 / 2} \pm \cos (\xi)\left(\beta^{\phi}+\hat{v}_{K}\right) \sigma_{0}^{1 / 2} \nonumber\\
& \left.-\frac{1}{4} \frac{\left(1+\sigma_{0}\right)^{1 / 2} \mp \cos (\xi )\hat{v}_{K} \sigma_{0}^{1 / 2}}{\hat{\gamma}_{K}^{2}\left(1+\sigma_{0}-\cos ^{2} (\xi) \hat{v}_{K}^{2} \sigma_{0}\right)}\right]
\end{align}
where we have assumed that the plasma is relativistic hot with $p=w/4$. The only difference between Eq.(\ref{ep}) and that of in the Kerr black hole case (see \cite{mr11}) is geometry quantities which are now substituted by metric (\ref{metric}).

Similar to the Penrose process \cite{p1,p2}, where energy extraction from rotating black holes occurs when specific circumstances are met, the same is true for magnetic reconnection, where the following requirements must be met
\begin{equation}\label{cd}
\epsilon_{-}^{\infty}<0 ,\quad
\Delta\epsilon_{+}^{\infty}=\epsilon_{+}^{\infty}-\left(1-\frac{\Gamma}{4(\Gamma-1)} \right)>0    
\end{equation}
for a relativistic hot plasma, i.e, $\Gamma=4/3$, we have $\Delta\epsilon_{+}^{\infty}=\epsilon_{+}^{\infty}$. As a result, black hole energy can only be extracted if the accelerated part of the plasma in a magnetic reconnection process acquires energy at infinity that is greater than its rest mass and thermal energies and the decelerated part of the plasma acquires negative energy as measured at infinity.

\section{Parameter spaces for energy extraction }\label{sec4}

The plasma energy at infinity density (\ref{ep}) depend on several critical parameters: the black hole mass $M$, the black hole spin $a$, the dominant reconnection radial location $r$, the plasma magnetization $\sigma_0$, the orientation angle $\xi$, the hairy parameters $\lambda$ and $h_0$. For simplicity, we will choose units by setting $M = 1$ in the rest of this paper, then $a$, $r$ and $h_0$ are measured in the unit of $M$. $\lambda$, $\sigma_0$ and $\xi$ are dimensionless parameters.

We will now demonstrate that, in a sizable range of parameter spaces, Comisso-Asenjo mechanism is a viable process to extract energy from hairy rotating black holes. In order to examine the effects of both hairy parameters $\lambda$ and $h_0$ on the magnetic reconnection and simplify our discussion, we choose four different combinations of hairy parameter in the following analysis, which take two typical values for each hairy parameter in its dominate range and permute them. As a result, we will use the following combinations: $(\lambda=0.5, h_0=0.5)$, $(\lambda=0.5, h_0=1.5)$, $(\lambda=1, h_0=1.5)$ and $(\lambda=1, h_0=0.5)$.

\begin{figure}[htbp]
  \includegraphics[scale=0.6]{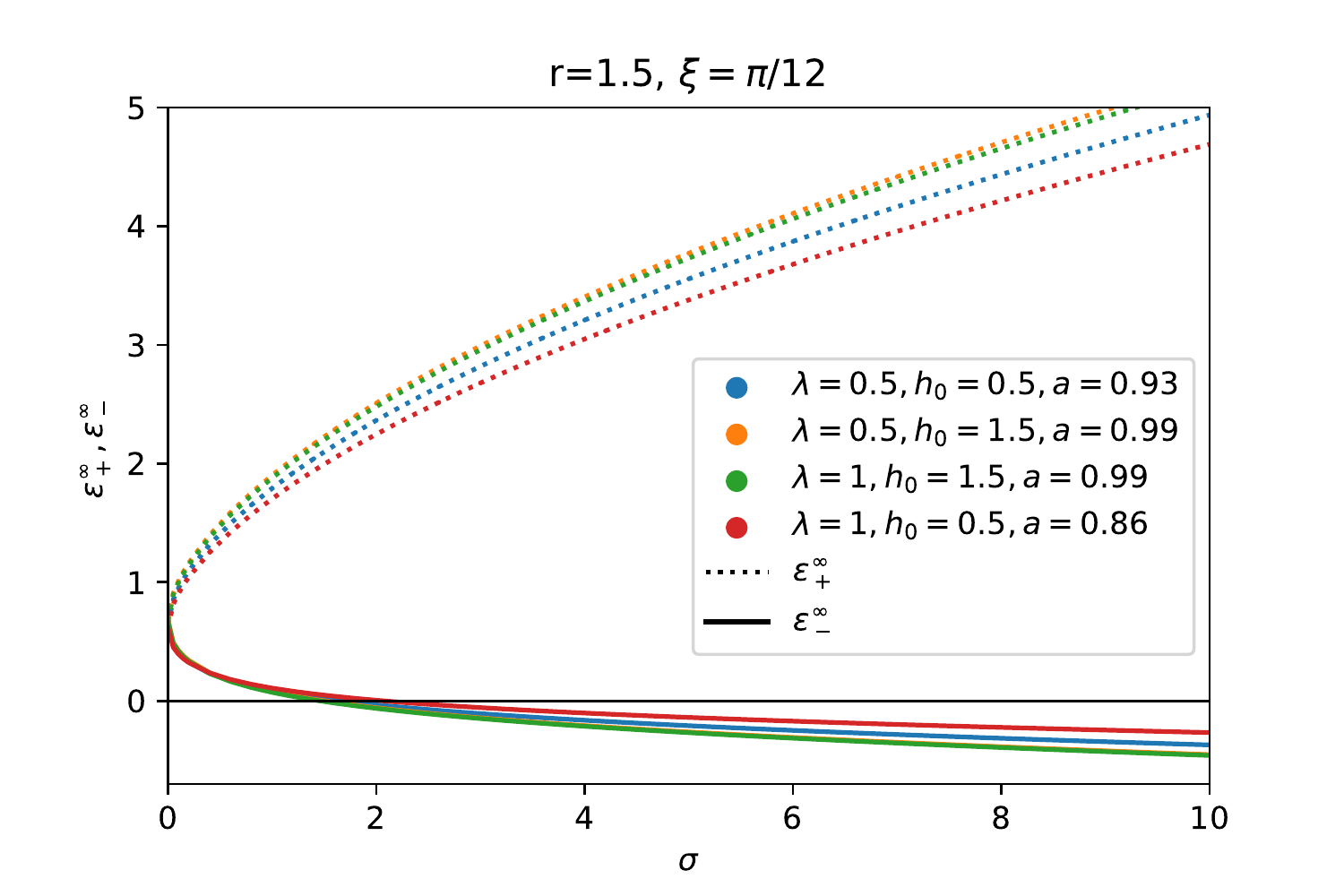}
\caption{The behaviors of $\epsilon_{+}^{\infty}$ (dotted curve) and $\epsilon_{-}^{\infty}$ (solid curve) as a function of plasma magnetization $\sigma_0\in[0, 10]$ with four different combinations of hairy parameter and black hole spin. The dominant reconnection radial location is taken as $r = 1.5$, with orientation angle $\xi=\pi/12$. The black solid line is $\epsilon_{+}^{\infty}=\epsilon_{-}^{\infty}$= 0 as reference}
\label{es}
\end{figure}

We first show the behavior of $\epsilon_{+}^{\infty}$ and $\epsilon_{-}^{\infty}$ as a function of plasma magnetization $\sigma_0$ in Fig.\ref{es}, taking $r=1.5$ and $\xi=\pi/12$ with four different combinations of hairy parameter and its corresponding maximal black hole spin for which it has two horizons. We can see that $\epsilon_{+}^{\infty}$ increases along with $\sigma_0$ while the $\epsilon_{-}^{\infty}$ decreases. In the entire range, $\epsilon_{+}^{\infty}>0$ and $\epsilon_{-}^{\infty}<0$ when $\sigma_0\gtrsim2$ for all the combinations of hairy parameter, which means it is easy to satisfy the conditions (\ref{cd}) and is very effective to extract energy from the hairy black hole via magnetic reconnection in sufficient parameter range. Although the difference of $\epsilon_{+}^{\infty}$ and $\epsilon_{-}^{\infty}$ between different combinations of hairy parameter are small, there are still notable information: the bigger the maximal black hole spin which the hairy parameter combination corresponds to, the larger $\epsilon_{+}^{\infty}$ and smaller $\epsilon_{-}^{\infty}$, almost the same when their maximal black hole spin are equal. That is to say that extract energy via the magnetic reconnection process has a positive correlation with the black hole spin. It may be easy to understand this since the higher black hole spin means more rotational energy. In addition, we also find that the difference of $\epsilon_{+}^{\infty}$ and $\epsilon_{-}^{\infty}$ between different $\lambda$ almost disappear when $h_0=1.5$, while the difference become notable when $h_0=0.5$. We can conclude that in order to let the effect of the hairy parameter $\lambda$ be significant, $h_0$ should be sufficiently small.

\begin{figure}[htbp]
  \includegraphics[scale=0.55]{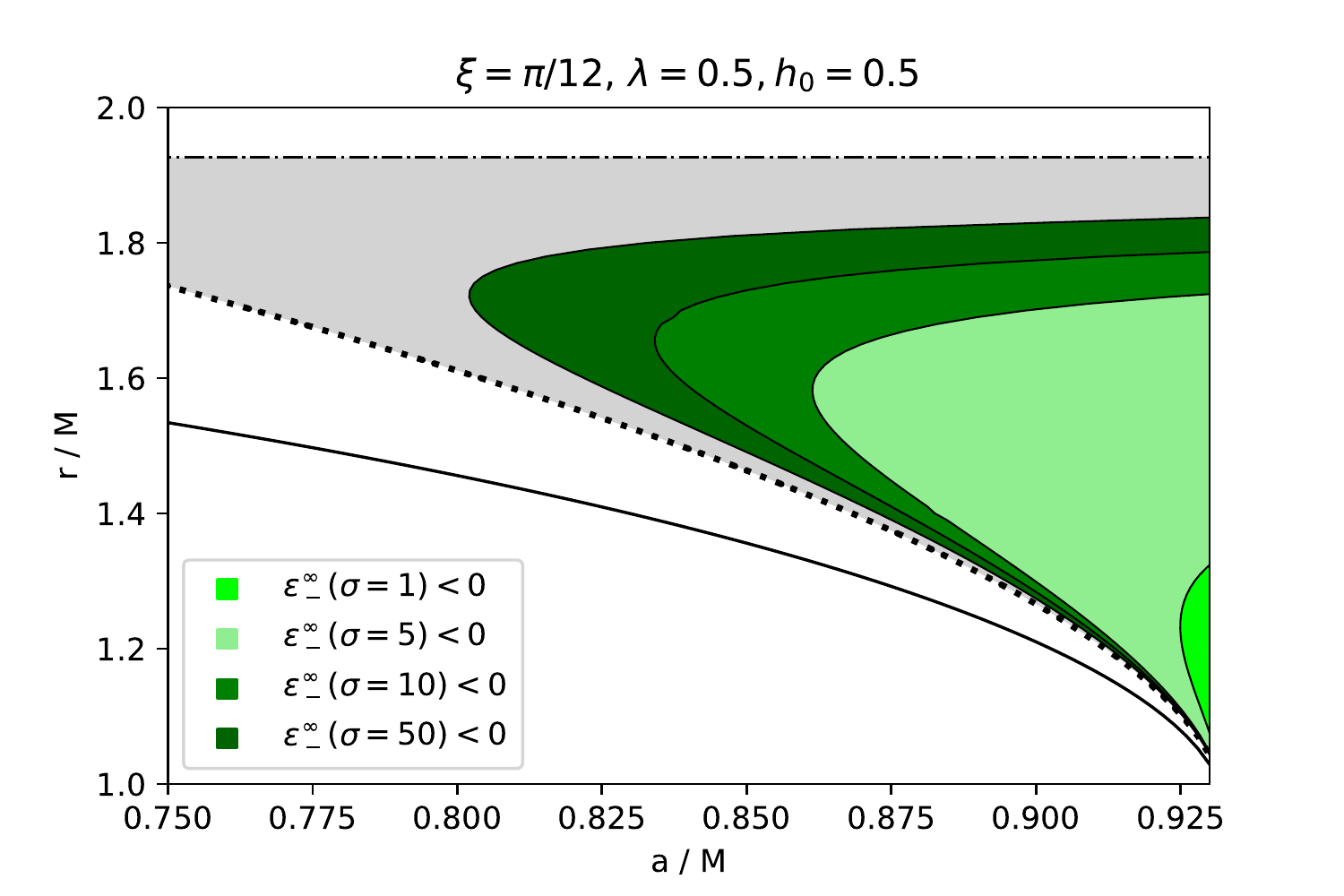}
  \includegraphics[scale=0.55]{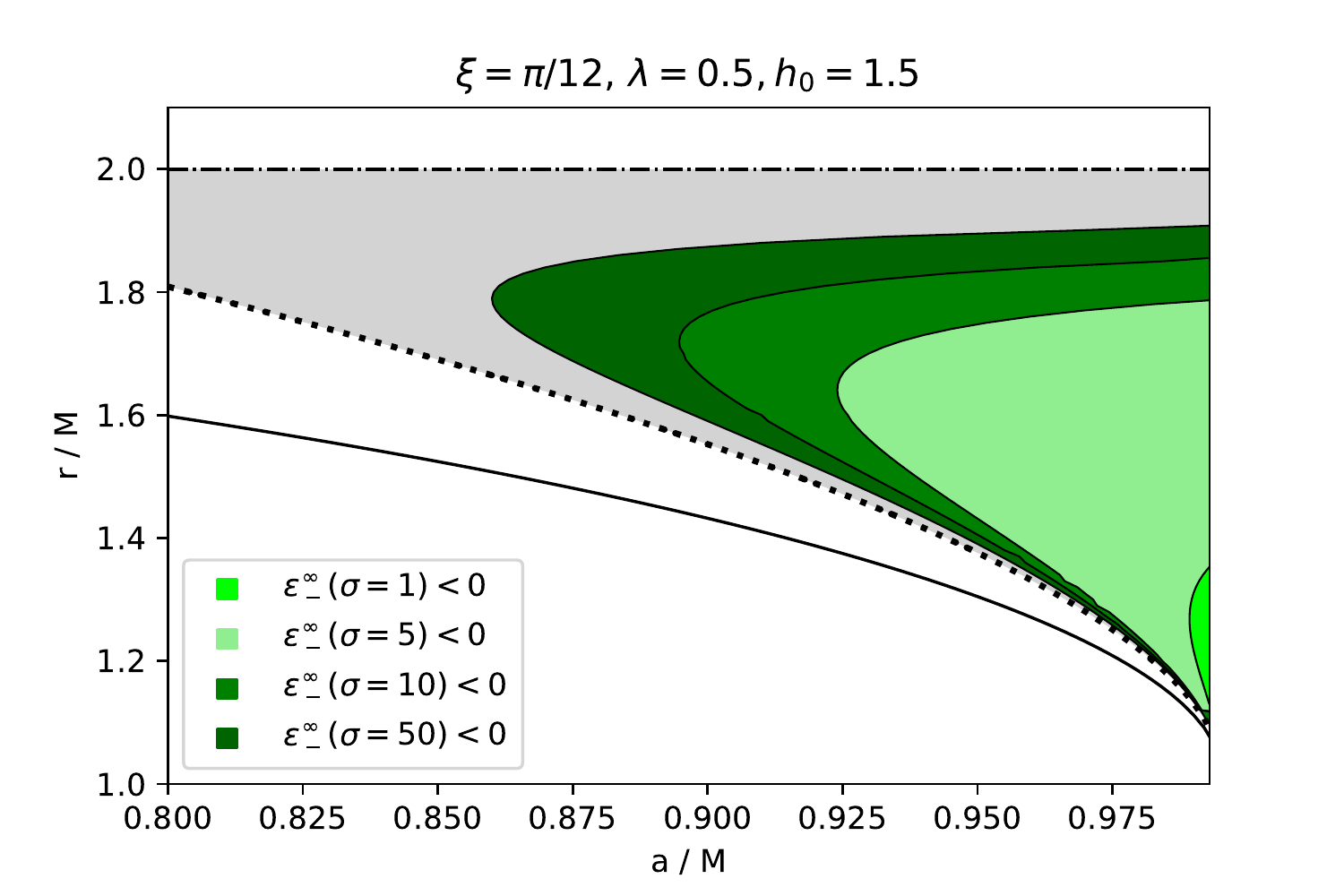}
  \includegraphics[scale=0.55]{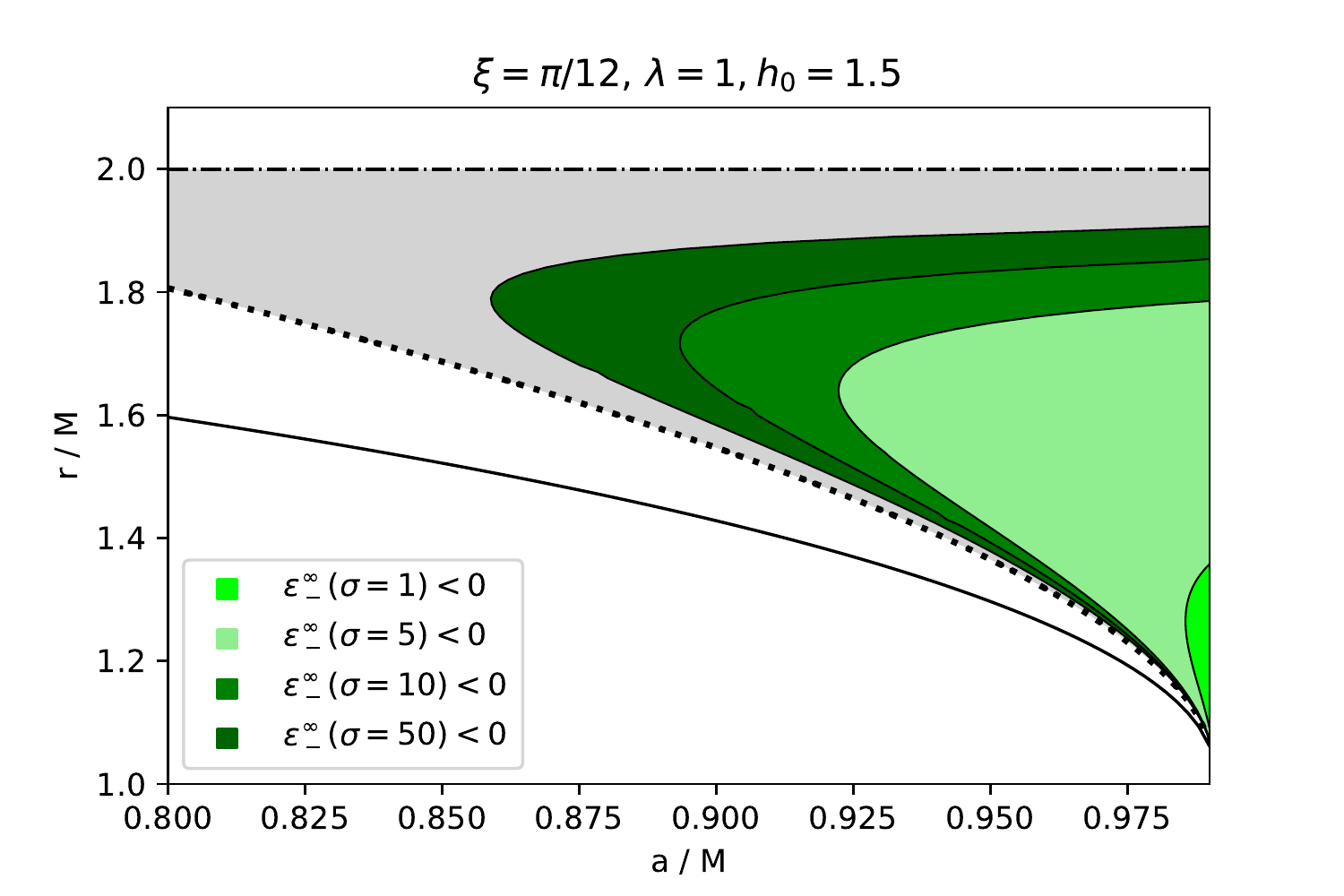}
  \includegraphics[scale=0.55]{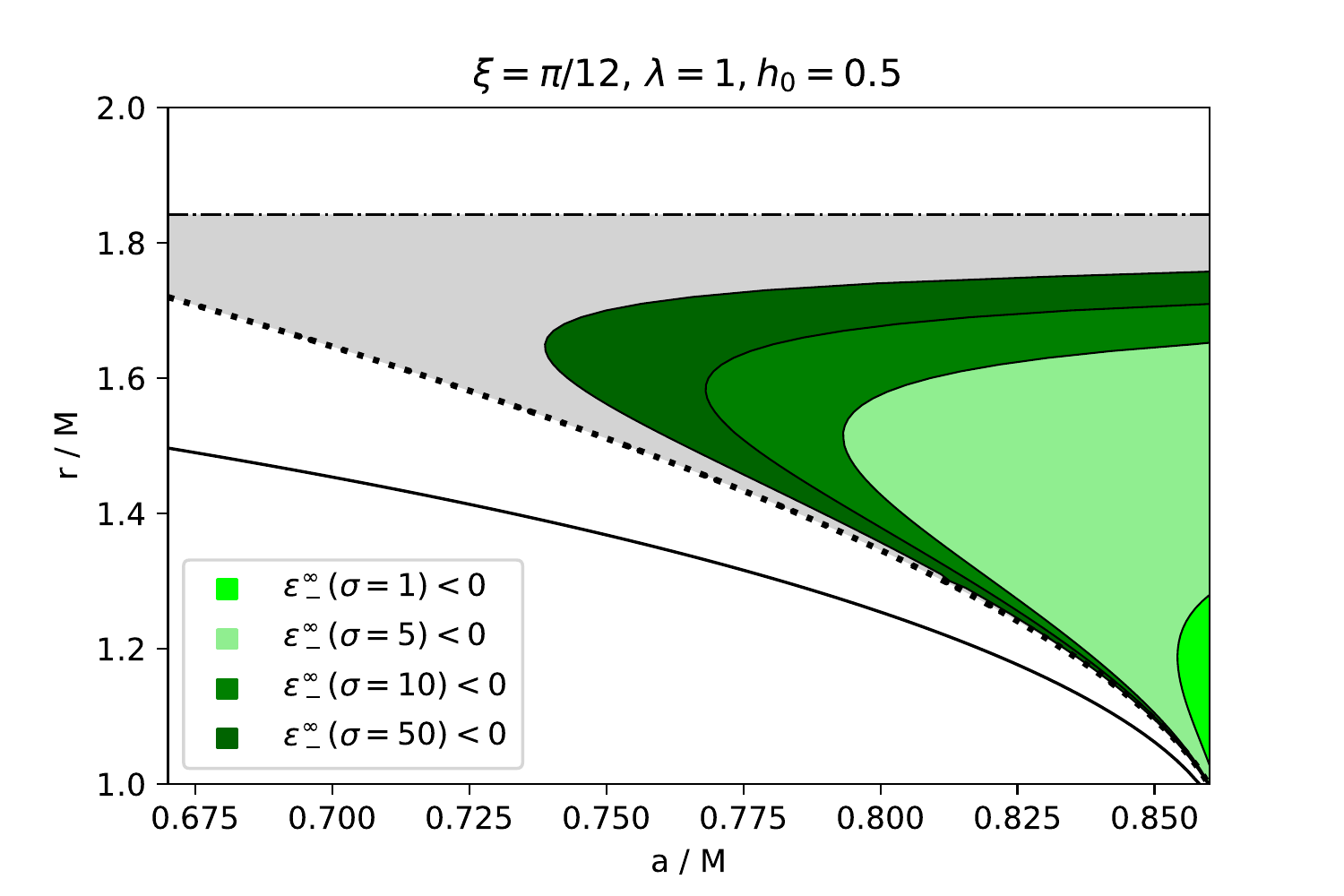}
\caption{The parameter space $(r -a)$ with $\xi=\pi/12$ and four different different combinations of hairy parameter. The colored regions are $\epsilon_{-}^{\infty}<0$ with $\sigma_0= 1, 5, 10, 50$. The gray area is the region where $\epsilon_{+}^{\infty}>0$. Black solid curves, black dotted curves, and black dot dashed curves are the radii of the outer event horizon, light ring, and outer ergosphere respectively.}
\label{ras}
\end{figure}

\begin{figure}[htbp]
  \includegraphics[scale=0.6]{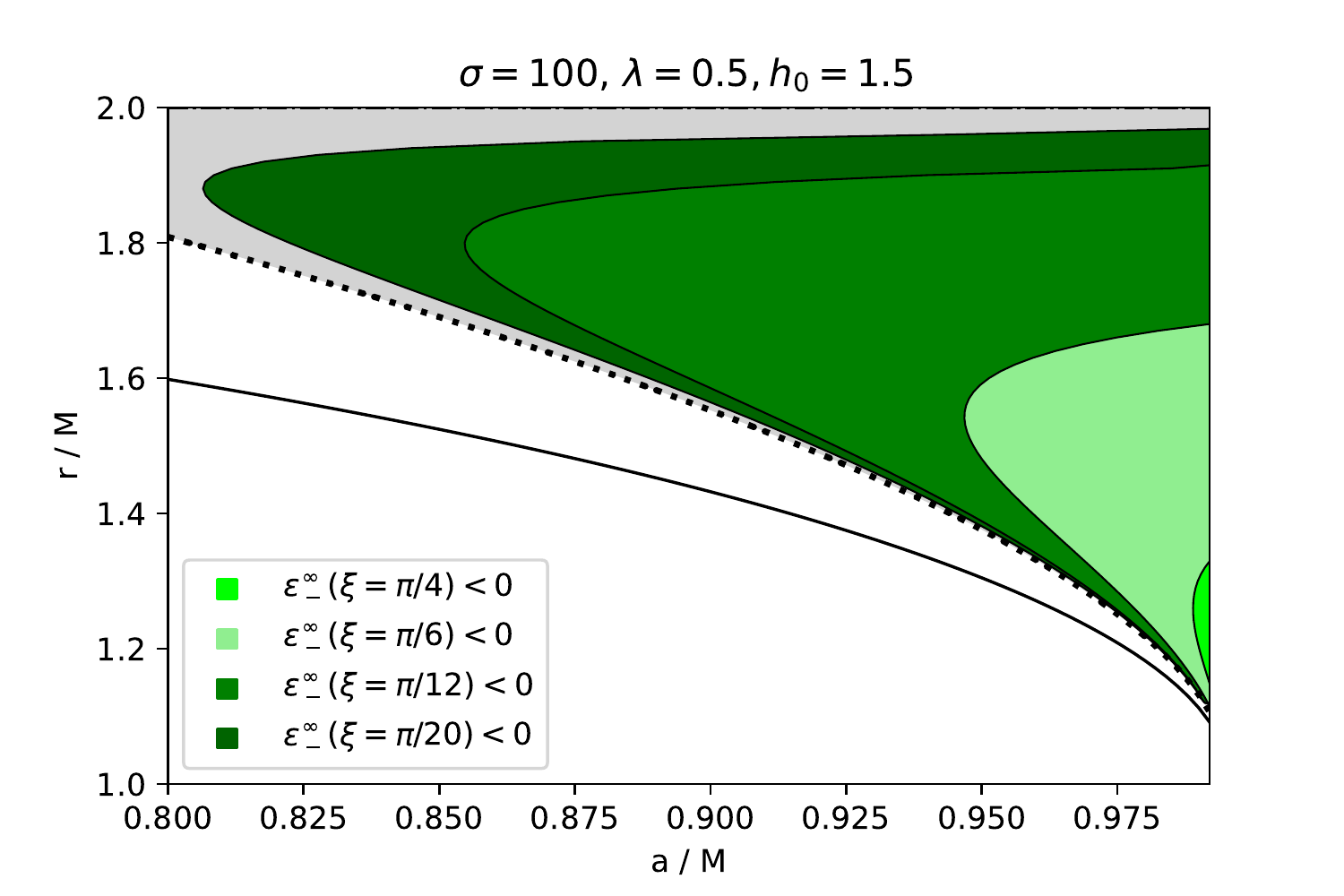}
\caption{The parameter space $(r -a)$ with $\sigma_0 = 100$, $\lambda=0.5$, $h_0=1.5$ and different orientation angle $\xi$. The colored regions are $\epsilon_{-}^{\infty}<0$ with $\xi = \pi/20, \pi/12, \pi/6, \pi/4$ respectively. The gray area is the region where $\epsilon_{+}^{\infty}>0$. Black solid curves, black dotted curves, and black dot dashed curves are the radii of the outer event horizon, light ring, and outer ergosphere respectively.}
\label{rax}
\end{figure}

To better understand how the black hole spin affect the energy extraction via magnetic reconnection, we plot the two dimensional $(r-a)$ parameter space in Fig.\ref{ras} and Fig.\ref{rax}. For the four subplots of Fig.\ref{ras}, they correspond to four combinations of hairy parameter. The maximal black hole spin, under which the rotating hairy black hole has two horizons, are different. We found that the smaller black hole spin require a relatively large radial location $r$ such that the energy extraction via the magnetic reconnection mechanism works. As we discussed before, here the parameter spaces for $(\lambda=0.5, h_0=1.5)$ and $(\lambda=1, h_0=1.5)$ is almost the same, this prove again that the effects of $\lambda$ depend how small the value of $h_0$. When $h_0=0.5$, $\lambda$ indeed cause a difference in the parameter space as well as the radii of the outer event horizon, light ring and outer ergosphere. In Fig.\ref{rax}, we aim to exam the effects of orientation angle $\xi$ on the $(r-a)$ parameter space by taking $\sigma_0 = 100$, $\lambda=0.5$, $h_0=1.5$ . We can see that the smaller orientation angle $\xi$, the larger region allowing to extract energy from the rotating hairy black hole via magnetic reconnection mechanism.

\section{Energy extraction power and efficiency}\label{sec5}

Black hole evolution and its associated astrophysical phenomenon depend significantly on the power and efficiency of energy extraction via Comisso-Asenjo mechanism. We will examine the energy extraction power and efficiency form the rotating hairy black hole. Comisso and Asenjo have proposed that \cite{mr11} these two quantities mostly depend on the rate at which plasma with negative energy-at-infinity density is absorbed. We denote the power as $P_{etr}$ and it can be well estimated by \cite{mr11}
\begin{equation}
P_{etr}=-\epsilon_{-}^{\infty}w_0 A_{in} U_{in}
\end{equation}
where $U_{in}=\mathcal{O}(10^{-1})$ and $\mathcal{O}(10^{-2})$ respectively refer to the collisionless and collisional regimes. $A_{in}$ is the cross-sectional area of the inflowing plasma, which can be estimated as $A_{in} \sim  r_E^2- r_{ph}^2 $ for rapid rotating black holes, with $r_E$ and $r_{ph}$ are the outer ergosphere and light ring of the black hole respectively.

\begin{figure}[htbp]
  \includegraphics[scale=0.6]{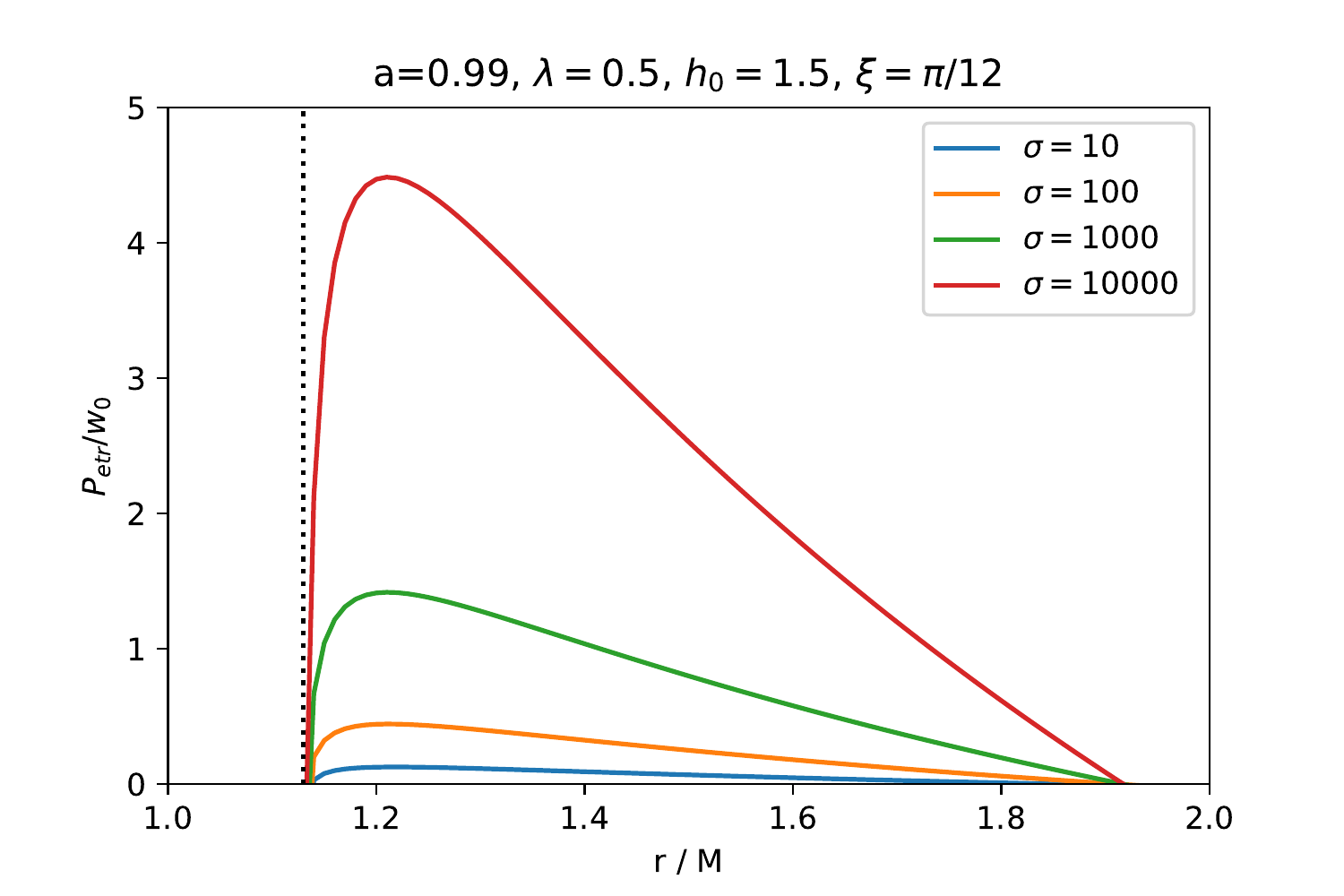}
\caption{$P_{etr}/w_0$ as a function of the dominant reconnection radial location $r$ with different plasma magnetization $\sigma_0 = 10, 100, 1000, 10000$, by taking $a = 0.99, \xi = \pi/12, \lambda = 0.5, h_0=1.5$ and $U_{in} = 0.1$. The vertical dotted line indicates the limiting circular orbit, i.e., light ring $r_{ph}$.}
\label{power}
\end{figure}

We first plot the ratio $P_{etr}/w_0$ as a function of the dominant reconnection radial location $r$ in Fig.\ref{power}, with different plasma magnetization $\sigma_0 = 10, 100, 1000, 10000$, by taking $a = 0.99$, $\xi = \pi/12$, $\lambda = 0.5$, $h_0=1.5$ and $U_{in} = 0.1$. The power extracted from the black hole rises monotonically as plasma magnetization rises. The power steadily declines after reaching a maximum at those locations close to the outermost circular orbit or light ring.

\begin{figure}[htbp]
  \includegraphics[scale=0.6]{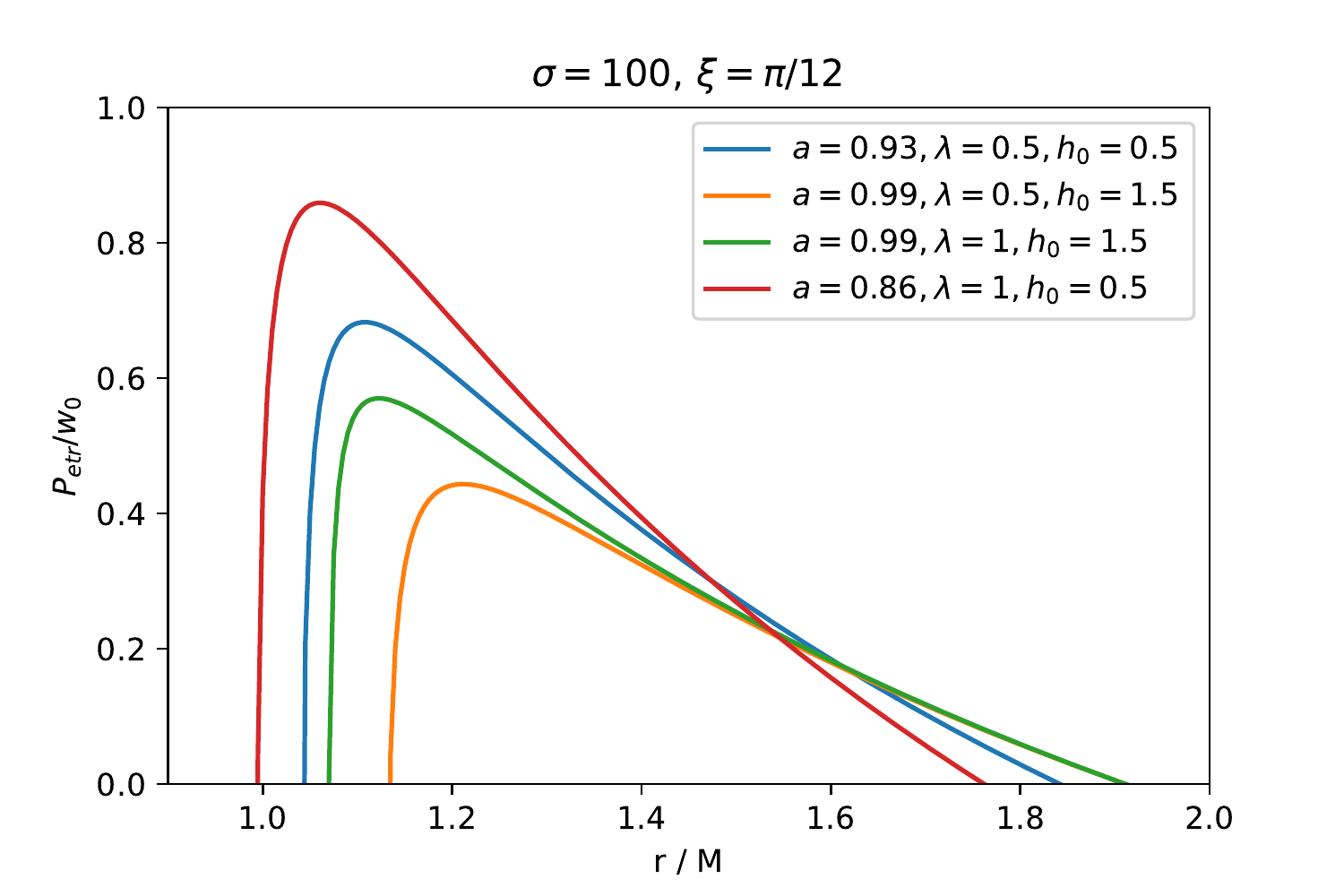}
\caption{$P_{etr}/w_0$ as a function of the dominant reconnection radial location $r$ with different combinations of the hairy parameter, by taking $\sigma=100, \xi = \pi/12$ and $U_{in} = 0.1$.}
\label{power-k}
\end{figure}

In order to examine the effects of the hairy parameters $\lambda$ and $h_0$, we plot the ratio $P_{etr}/w_0$ as a function of the dominant reconnection radial location $r$ in Fig.\ref{power-k}, but with different combinations of hairy parameter, taking $\sigma=100, \xi = \pi/12$ and $U_{in} = 0.1$. The results show that they all rise sharply when the location closer to their light ring and then decline along with the increases in radial location $r$. The higher the peaks, the faster the decline. The critical locations $r$ where $P_{etr}/w_0=0$ and the peaks where $P_{etr}/w_0$ reach the maximum also shift with the hairy parameters. The bigger value of $\lambda$, the smaller critical locations and the higher peaks when $h_0$ keep unchanged. The bigger value of $h_0$, the bigger critical locations and the lower peaks when $\lambda$ keep unchanged.

Next, we assess the efficiency. It is practical to define the efficiency as \cite{mr11}
\begin{equation}
\eta=\frac{\epsilon_{+}^{\infty}}{\epsilon_{+}^{\infty}+\epsilon_{-}^{\infty}}
\end{equation}
if $\eta >1$, then the energy will be extracted from the rotating hairy black hole. 

\begin{figure}[htbp]
\includegraphics[scale=0.6]{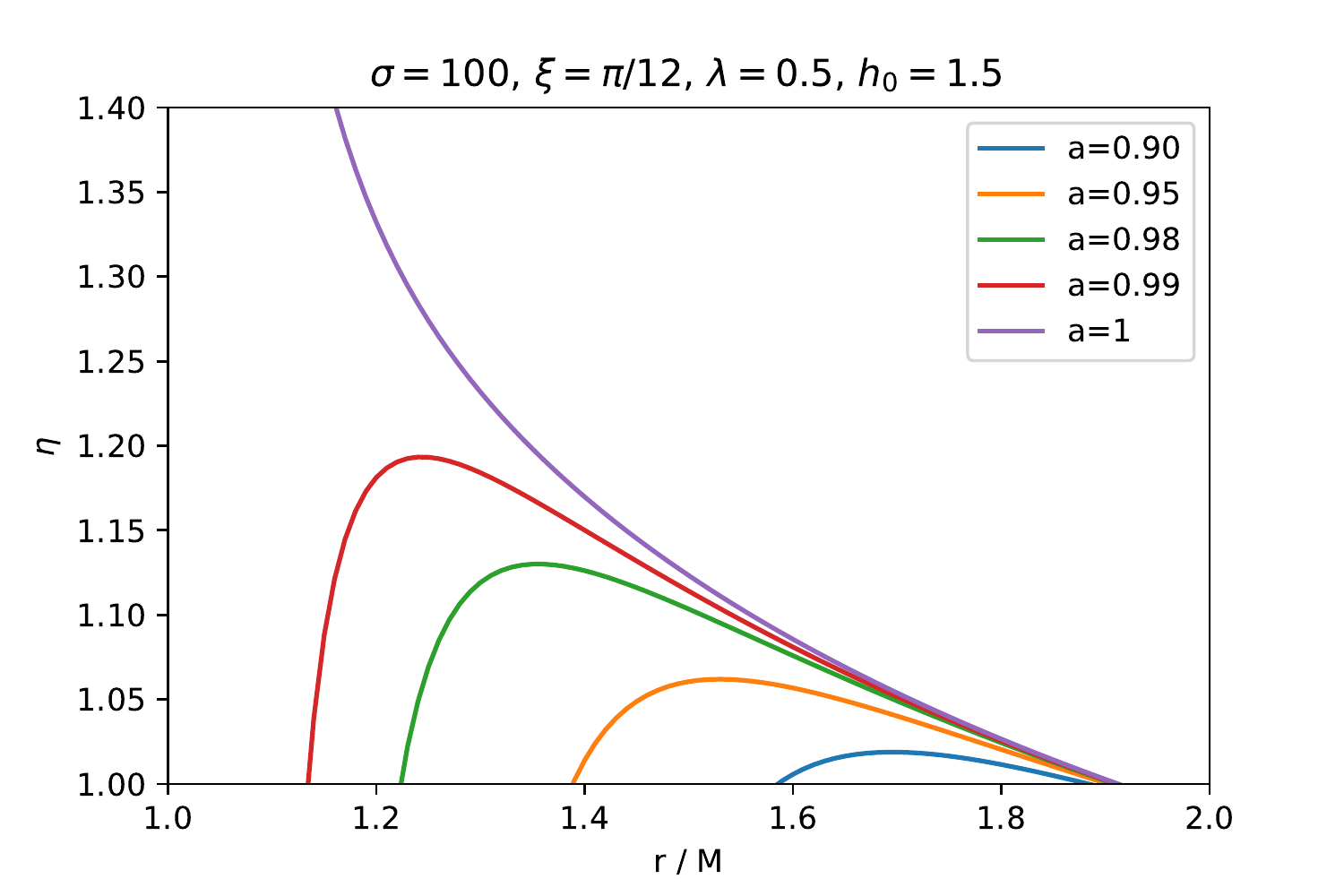}
\caption{Efficiency $\eta$ of the magnetic reconnection process as a function of the dominant reconnection radial location $r$ with different black hole spin $a = 0.90, 0.95, 0.98, 0.99, 1$, taking $\sigma=100, \xi=\pi/12, \lambda=0.5, h_0=1.5$. Note that the $a=1$ case is the extreme black hole.}
\label{eta}
\end{figure}

We plot the efficiency $\eta$ as a function of the dominant reconnection radial location $r$ in Fig.\ref{eta} with different black hole spin $a = 0.90, 0.95, 0.98, 0.99, 1$, taking $\sigma=100, \xi=\pi/12, \lambda=0.5, h_0=1.5$. Note that the $a=1$ case is the over-extreme black hole. With regard to the non-extreme situation, Fig.\ref{eta} shows that the efficiency sharply increases with location $r$ that are closer to the light ring and gradually declines after the peaks. As the black hole spin increase, the peaks shift to smaller locations and grow to higher values.

\begin{figure}[htbp]
  \includegraphics[scale=0.6]{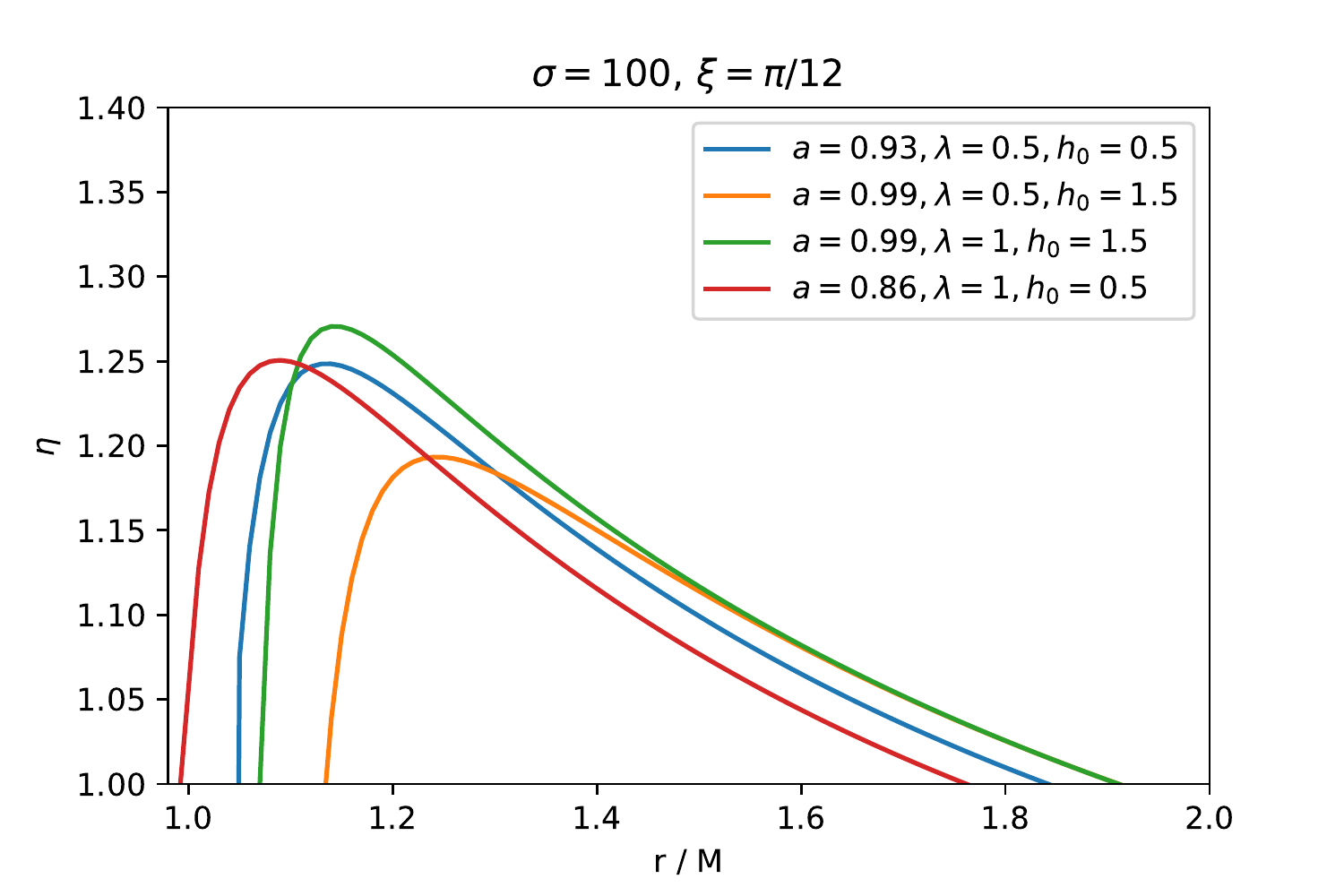}
\caption{Efficiency $\eta$ of the magnetic reconnection process as a function of the dominant reconnection radial location $r$  with different combinations of hairy parameter, taking $\sigma=100, \xi=\pi/12$.}
\label{eta-k}
\end{figure}

In addition, we plot the efficiency $\eta$ in Fig.\ref{eta-k} as a function of the dominant reconnection radial location $r$ with different combinations of hairy parameter, taking $\sigma=100, \xi=\pi/12$. There are also peaks near their light ring, and then all $\eta$ decrease along with the radial location $r$, which seems that the decline rate are the same at sufficient large location $(r\approx 1.4)$ regardless the hairy parameter combinations. However, the position and value of the peaks depend on the  combinations of hairy parameter.

\begin{figure}[htbp]
  \includegraphics[scale=0.6]{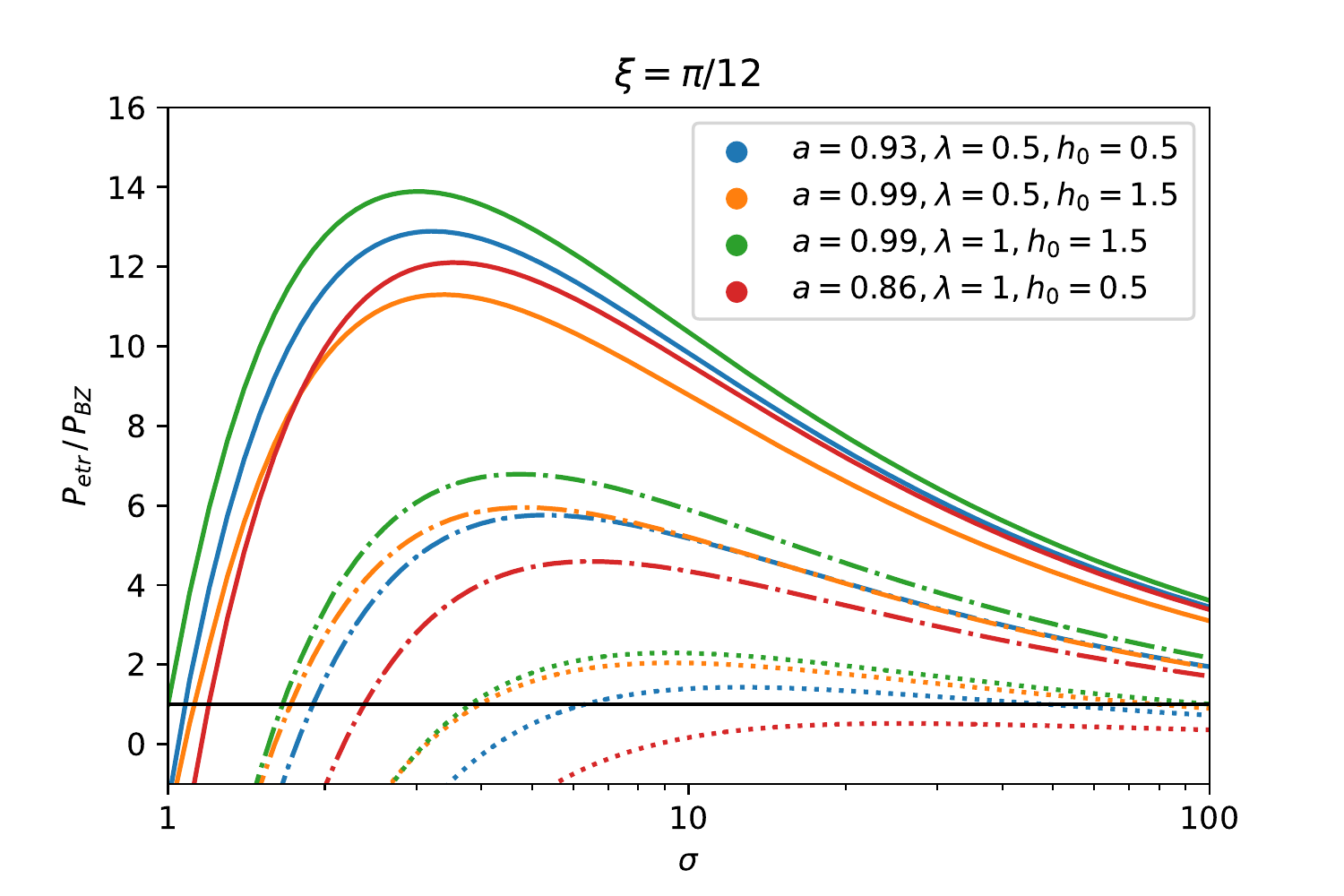}
\caption{Power ratio $P_{etr}/P_{BZ}$ as a function of the plasma magnetization $\sigma_0$ with different dominant reconnection radial location $r = 1.3$ (solid curve), $1.5$ (dot-dashed curve), $1.7$ (dotted curve), by taking the orientation angle $\xi=\pi/12$. The coefficients are taken as $\kappa = 0.05$, $C_1 = 1.38$, $C_2 = -9.2$ respectively. The black solid line is $P_{etr}/P_{BZ} = 1$ as reference.}
\label{ratio}
\end{figure}

What's more, we would like to make a power comparison between the energy extraction via Comisso-Asenjo mechanism and Blandford-Znajek mechanism. Regarding to the Blandford-Znajek mechanism, it extract the rotation energy of black hole through a magnetic field that threads the event horizon. In the maximum efficiency conditions \cite{bz1,bz2,bz3}, the power of energy extraction is given by \cite{bz4,bz5,bz6}
\begin{equation}\label{bzz}
P_{BZ} =\frac{\kappa}{16\pi}\Phi_{H}^2(\Omega_{H}^2+C_1 \Omega_{H}^4+C_2\Omega_{H}^6+\mathcal{O}(\Omega_{H}^8))
\end{equation}
where $\kappa$ is a numerical constant related to the magnetic field configuration. $C_1$ and $C_2$ are the numerical coefficients. The magnetic flux crossing the black hole event horizon is given by $\Phi_{H}=\frac{1}{2} \int_{\theta}\int_{\phi}\left | B_r \right | \sqrt{-g} d\theta d\phi=2\pi(r_+^2+a^2)B_0 sin(\xi)$. The angular velocity at the event horizon is $\Omega_{H}=2 a r_+ (1-\frac{1}{2}\lambda r_+e^{-r_+/(1-h_0/2)})/{\left(r_+^{2}+a^{2}\right)^{2}}$. Since the fundamental magnetic field configurations are the same as in the Kerr case, we expect that the difference in the spacetime metric will only have an impact on the Blandford-Znajek process through geometry quantities. As a result, we just change the geometry quantities in the Blandford-Znajek power of Kerr black hole and obtained Eq.(\ref{bzz}). Then, the power ratio between these two mechanism is 
\begin{equation}\label{pr}
\frac{P_{etr}}{P_{BZ}}\sim \frac{-4\epsilon_{-}^{\infty} A_{in} U_{in}}{\pi \kappa \sigma_0(r_+^2+a^2)^2 sin^2(\xi) (\Omega_{H}^2+C_1 \Omega_{H}^4+C_2\Omega_{H}^6)}
\end{equation}
Taking the orientation angle $\xi=\pi/12$, the coefficients $\kappa = 0.05$, $C_1 = 1.38$ and $C_2 = -9.2$ \cite{mr11}, we show the power ratio $P_{etr}/P_{BZ}$ as a function of the plasma magnetization $\sigma_0$ in Fig.\ref{ratio} with different dominant reconnection radial location $r = 1.3, 1.5, 1.7$. As the plasma magnetization gets closer to the critical value, we can observe that all the power ratios rise along with it. Once their maximum values are reached, power ratios decrease along with the plasma magnetization. The distinct dependence on plasma magnetization is what causes the power ratios Eq.(\ref{pr}) to decline as plasma magnetization increases. Since we have $P_{etr}/P_{BZ} \sim 1/\sigma_0^{1/2}$ when $\sigma_0 \to \infty$, which drops as plasma magnetization increases. On the other hand, the force-free electrodynamics approximation \cite{bz4,bz5,bz6} of the Blandford-Znajek power is invalid when $\sigma_0 \sim 1$, therefore, the power ratios in this case can only be seen as an effective comparison. We also notice that the smaller the location $r$, the faster the rising and decline of the curves. The different combinations of hairy parameter only slightly change the curves comparing to the location $r$. Nevertheless, the Comisso-Asenjo mechanism is a very promising and significant energy extraction mechanism from rotating hairy black holes since the power ratios are greater than 1 throughout a very wide parameter range.

\section{conclusion}\label{sec6}

The rotating hairy black hole is a very encouraging black hole solution beyond classical general relativity. The existence of new hair originated from unknown fundamental field, such as dark matter and dark energy, will breakdown the No-Hair theorem of general relativity and lead to new theory for gravity as well as spacetime. Therefore, the phenomenological investigation of rotating hairy black hole spacetime is crucial for testing this hypothesis and laid foundation for observations. In this paper, we studied the energy extraction from rotating hairy black hole  (\ref{metric}) via magnetic reconnection process in the ergosphere. 

We started by presenting Comisso-Asenjo formulas, i.e., the plasma energy at infinity density (\ref{ep}) associated with the accelerated and decelerated parts as well as the conditions (\ref{cd}) allowing the energy extraction to occur. Then, all the following analysis are based on these. 

We first emphasize our discussion on how the plasma magnetization and hairy parameter will affect the behaviour of the plasma energy-at-infinity density per enthalpy $\epsilon_{+}^{\infty}$ and $\epsilon_{-}^{\infty}$, taking the orientation angle $\xi=\pi/12$ and dominant reconnection radial location $r=1.5$. The results show that $\epsilon_{+}^{\infty}$ increases monotonically along with plasma magnetization while $\epsilon_{-}^{\infty}$ decreases with it. In addition, we found that in order to make the effects of the hairy parameter $\lambda$ become significant, the other hairy parameter $h_0$ should be sufficiently small. The amount of energy extraction has a positive correlation with the maximal black hole spin which the hairy parameter combinations correspond to. Then, we studied the how the black hole spin affect the energy extraction by showing the two dimensional $(r-a)$ parameter space with different combinations of hairy parameter, taking orientation angle $\xi=\pi/12$. We found that the smaller the black hole spin the relatively larger radial location $r$ such that the energy extraction works.

We also investigated how the power and efficiency of the energy extraction through magnetic reconnection will be impacted by the hairy parameters as well as the other critical parameters. They all have peaks near their light ring and decrease with radial location after reaching the peaks. However, the location and value of the peaks depend on the combinations of hairy parameter. At the end, we compared the power ratio between the Comisso-Asenjo mechanism and Blandford-Znajek mechanism. It turns out that the energy extraction via Comisso-Asenjo mechanism is more effective in a sufficiently wide parameter range. Especially, the closer of the radial location to the event horizon, the greater the power ratio. The effects of different combinations of hairy parameter on the power ratios are small comparing to that of the dominant reconnection radial location.

\section*{acknowledgments}
Zhen Li would like to thank the DARK cosmology centre at Niels Bohr Institute for supporting this research. Zhen Li is also financially supported by the China Scholarship Council. Faqiang Yuan is supported by the National Natural Science Foundation of China (Grant Nos.12275022). We thank Filippo Camilloni for helpful discussions.  We also would like to thank Luca Comisso for his suggestions and comments which has significantly improved this paper.
\\
\\

\end{document}